\documentstyle[12pt,aaspp4]{article}
 
\begin{document}
 
\title{Possible Post-AGB Stars \footnote{Based on spectra obtained with
the Hobby-Eberly Telescope, which is a joint project of 
the University of Texas at Austin, the Pennsylvania State University, 
Stanford University, Ludwig-Maximillians-Universit\"at M\"unchen, and
Georg-August-Universit\"at G\"ottingen.}}
 
\author{G. Wallerstein} 
\affil{Astronomy Department, Box 351580, University of Washington, Seattle, WA 98195}
\author{G. Gonzalez}
\affil{Department of Physics and Astronomy, Iowa State University, Ames, IA 50011 }
\author{M.D. Shetrone} 
\affil{University of Texas, McDonald Observatory, Fort Davis, Tx 79734}

\begin{abstract}

Radial velocities were measured for 5 potential post-AGB stars, two in 
the globular cluster NGC 1851 and three in the dSph galaxy Ursa Minor.
All five potential PAGB stars were found to be radial velocity non-members. 

\end{abstract}
 
 
\section{The Spectra}

Eskridge \& Schweitzer 2001 reported proper motion
memberships for 3 stars that fall in the post-AGB region of the C-M diagram of
the UMi system.  In 2003 we obtained spectra during heavily overcast full moon
engineering time 
at the Hobby-Eberly Telescope (HET) with the High Resolution Spectrograph 
(HRS, Tull et al. 1998).  The spectra were taken in the lowest resolution mode
($R=15,000$) and had signal-to-noise ratios between 10 and
20 per resolution element.    The spectra where reduced with 
IRAF\altaffilmark{2}\altaffiltext{2}{IRAF is distributed by the National Optical
Astronomy Observatories, which are operated by the Association of Universities 
for Research in Astronomy, Inc., under contract to the National Science Foundation.} 
{\it echelle} scripts.  The velocities for the three Ursa Minor stars are given in Table 1.
Since the velocity of the UMi galaxy is known to be -248 km/sec, it
appears that these stars are not members.

In addition in 1997 the authors obtained 4-m echelle spectra of 2
likely post-AGB stars in the globular cluster NGC 1851, stars W266 and W464
 (Walker 1992). Both stars showed velocities near zero and hence are 
very likely to also be non-members of NGC 1851.

\begin{deluxetable}{ccccccc}
\tablecaption{Results from HET$+$HRS 
\label{table1}}
\tablewidth{0pt}
\tablehead{
\colhead{Star} &
\colhead{$\alpha$} &
\colhead{$\delta$} &
\colhead{V} &
\colhead{B-V} &
\colhead{P\tablenotemark{1}} &
\colhead{V$_r$ (km/s)} \\
}
\startdata

746  &15:08:51.6 &  +67:25:11 &16.17 &0.74&     0.51 &      -52  \\
783  &15:07:14.5 &  +67:20:30 &15.86 &0.71&     0.96 &      -29    \\
1406 &15:07:23.9 &  +67:11:05 &16.09 &1.05&     0.67 &      +13:   \\
 
\enddata
\tablenotetext{1}{ The proper motion probabilities taken from Eskridge \& Schweitzer 2001}
 
\end{deluxetable}
 
\end{document}